\documentclass[12pt,a4paper]{article}
\usepackage{epsfig}
\usepackage{amsmath}
\usepackage{amsfonts}
\usepackage{amssymb}
\usepackage{graphicx}
\usepackage[T1]{fontenc}
\usepackage[utf8]{inputenc}
\usepackage{authblk}

\input{tcilatex}
\begin{document}

\title{Inverted oscillator: pseudo hermiticity and coherent states }
\author{ Rahma Zerimeche$^{a,b}$, Rostom Moufok$^{a}$, Nadjat Amaouche$^{a}$
and Mustapha Maamache$^{a}$\thanks{%
E-mail: maamache@univ-setif.dz} \\
$^{a}$Laboratoire de Physique Quantique et Syst\`{e}mes Dynamiques,\\
Facult\'{e} des Sciences, Ferhat Abbas S\'{e}tif 1, S\'{e}tif 19000, Algeria.%
\\
$^{b\text{ \ }}$Physics Department, University of Jijel, BP 98, Ouled Aissa,
18000 Jijel, Algeria{\small .}\\
}
\date{}
\maketitle

\begin{abstract}
It is known that the standard and the inverted harmonic oscillator are
different. Replacing thus\textrm{\ }$\omega $\textrm{\ }by\textrm{\ }$\pm
i\omega $\textrm{\ }in the regular oscillator is necessary going to give the
inverted oscillator\textrm{\ }$H^{r}$\textrm{. }This replacement would lead
to anti- $\mathcal{PT}$-symmetric harmonic oscillator Hamiltonian\textrm{\ }$%
\left( \mp iH^{os}\right) $\textrm{. }The pseudo-hermiticity relation has
been used here to relate the anti-$\mathcal{PT}$-symmetric harmonic
Hamiltonian to the inverted oscillator. By using a simple algebra, we
introduce the ladder operators describing the inverted harmonic oscillator
to reproduce the analytical solutions.We construct the inverted coherent
states which minimize the quantum mechanical uncertainty between the
position and the momentum.

\textit{This paper is dedicated to the memory of Omar Djemli and Nouredinne
Mebarki who died due to covid 19 }
\end{abstract}

\section{\protect\bigskip Introduction}

The inverted oscillator, equipped with a potential exerting a repulsive
force on a particle, has been widely studied\ \cite{Barton}-\cite{Bha}. Such
system can be completely solved as the standard harmonic oscillator whose
properties are well known.

However, the physics of the inverted harmonic oscillator is different,
because its energy spectrum is continuous and its eigenstates are no longer
square integrable.The inverted oscillator can be applied to various physical
systems such as \cite{Barton},\ \cite{cho}-\cite{tar}, the tunneling
effects, the mechanism of matter-wave bright solitons, the cosmological
model, and the quantum theory of measurement.

In fact, the predominant idea in the literature is that the inverted
oscillator is obtainable from the harmonic oscillator by the replacement $%
\omega \rightarrow \pm i\omega $. Of course, in spite of many useful
analogies, it is important to know that the two oscillators (harmonic and
inverted) reveal different characteristics. In other words, the inverted
oscillator generates a wave packet which are not square integrable and there
is no zero-point energy. In comparison with the harmonic oscillator, the
physical applications of the inverted harmonic oscillators are limited,
since their Hamiltonian is parabolic and the eigenstates are scattering
states. The analytic continuation of angular velocity $\omega \rightarrow
\pm i\omega $ performs a transformation of a non-Hermitian harmonic
oscillator $(\mp iH^{os})$ to inverted one $H^{r}.$

In general, non-Hermitian Hamiltonians have been used to describe several
physical dissipative systems. Such Hamiltonians do not cause a legitimate
probabilistic interpretation due to the shortage of the unitarity condition
in their corresponding quantum description.\ In non-Hermitian quantum
mechanics it, was found that the criteria for a quantum Hamiltonian to have
a real spectrum is that it possesses an unbroken $\mathcal{PT}$symmetry ($%
\mathcal{P}$ is the space-reflection operator or parity operator, and $%
\mathcal{T}$ is the time-reversal operator) \cite{Bender98,Bender2002}.\ The
concept of $\mathcal{PT}$-symmetry has found applications in several areas
of physics. Once the non-Hermitian Hamiltonian $H$ is invariant under the
combined action of $\mathcal{PT}$ (i.e. $H$ commutes with $\mathcal{PT}$)
and its eigenvectors are also those of the $\mathcal{PT}$ operator, then the
energy eigenvalues $E$ of the system are real and in this case the $\mathcal{%
PT}$-symmetry is unbroken.

An alternative approach to explore the basic structure responsible for the
reality of the spectrum of a non- Hermitian Hamiltonian is by the notion of
the pseudo-hermiticity introduced in Ref. \cite{Mostafa1}.\ An operator $H$
is said to be pseudo-Hermitian if \ 
\begin{equation}
H^{\dagger }=\eta H\eta ^{-1},  \label{1}
\end{equation}%
where the metric operator 
\begin{equation}
\eta =\rho ^{\dagger }\rho ,\text{ }\eta ^{-1}=\left( \rho ^{\dagger }\rho
\right) ^{-1},
\end{equation}%
is a linear, invertible and Hermitian operator, we say that the Hamiltonian
is pseudo-Hermitian or quasi-Hermitian if it satisfies the relation (\ref{1}%
).

The pseudo-Hermiticity allows to link the pseudo-Hermitian Hamiltonian $H$
with an equivalent Hermitian Hamiltonian $h$ 
\begin{equation}
h=\rho H\rho ^{-1},  \label{ph}
\end{equation}%
where the operator $\rho $ called Dyson operator is linear and invertible.\
Due to the energy spectrum of $(\pm iH^{os})$ being completely imaginary, we
notice that $(\mp iH^{os})$ is anti-$\mathcal{PT}$-symmetric\ i.e.%
\begin{equation}
\mathcal{PT}(\pm iH^{os})\mathcal{PT}=(\mp iH^{os}).
\end{equation}

We recall that a $\mathcal{PT}$-symmetric system can be transformed to an
anti-$\mathcal{PT}$-symmetric one by replacing $H^{os}\rightarrow (\pm
iH^{os})$ \cite{Ge}-\cite{Peng}, which changes the physical structure of the
system. In other words, a Hamiltonian $H$ is said to be anti-$\mathcal{PT}$%
-symmetric if it anticommutes with the $\mathcal{PT}$ operator $\left\{ 
\mathcal{PT},H\right\} =0$. In analogy with the $\mathcal{PT}$-symmetric
case, we call the anti-$\mathcal{PT}$-symmetry of Hamiltonian $H$ unbroken
if all of the eigenfunctions of $H$ are eigenfunctions of $\mathcal{PT}$,
i.e. when the energy spectrum of $H$ is entirely imaginary $E$ $=iE^{\ast }$ 
\cite{Kheniche}.

In this paper, we generate from the anti-$\mathcal{PT}$-symmetric
Hamiltonian\ $(\pm iH^{os})$ an inverted Hermitian harmonic oscillator-type $%
H^{r}$ and also its solution. In Section 2, we recall briefly some
properties of the standard harmonic and inverted oscillators In Section 3,
introducing an appropriate quantum metric, we link the anti-$\mathcal{PT}$%
-symmetric Hamiltonian \ $(\pm iH^{os})$ to the inverted oscillator
Hamiltonian $H^{r}.$This procedure allows us to obtain the ladder operators,
the set of solutions and also to define the full orthonormalization relation
of the eigenstates for inverted harmonic oscillator $H^{r}$. In Section 4,
using the ladder operators, we will address the problem constructing of
coherent states associated to inverted oscillator\ $H^{r}$. We obtain the
mean values of the position and momentum operators in the evolved coherent
states and furthermore we calculate the corresponding Heisenberg
uncertainty. An outlook over the main results is given in the conclusion.

\section{Summary of standard harmonic and the inverted oscillators}

Let us recall briefly the ladder operator approach of the usual harmonic
oscillator: 
\begin{equation}
H^{os}=\frac{1}{2m}p^{2}+\frac{1}{2}m\omega ^{2}x^{2}=\frac{\hbar \omega }{2}%
\left( a^{+}a+aa^{+}\right) ,  \label{Hosc}
\end{equation}%
where

\begin{equation}
a=\sqrt{\frac{m\omega }{2\hbar }}x+i\frac{p}{\sqrt{2m\hbar \omega }}\text{ \
\ \ \ ,\ \ \ \ \ }a^{\dagger }=\sqrt{\frac{m\omega }{2\hbar }}x-i\frac{p}{%
\sqrt{2m\hbar \omega }},  \label{L1}
\end{equation}%
The operators $a$ and $a^{+}$ satisfying the commutation relation%
\begin{equation}
\left[ a,a^{\dagger }\right] =1.
\end{equation}

Were introduced to facilitate the solution of the eigenvalue problem.
Eigenstates of (\ref{Hosc}) in Fock space are the Fock or number states $%
\left\vert n\right\rangle ^{os}$ with the eigenvalues $\omega \left(
n+1/2)\right) ,$ where $a\left\vert n\right\rangle ^{os}=\sqrt{n}\left\vert
n-1\right\rangle ^{os},$ $a^{\dagger }\left\vert n\right\rangle ^{os}=\sqrt{%
n+1}\left\vert n+1\right\rangle ^{os}$ and $n$ is a non-negative integer.

We then have a nice mechanism for computing the eigenstates of the
Hamiltonian, but we can also express expectation values using the raising
and lowering operators. This leads to the useful representation of $x$ and $%
p $ : 
\begin{equation}
x=\sqrt{\frac{\hbar }{2\omega m}}\left( a^{\dagger }+a\right) \text{ \ \ \ \
\ \ \ \ , \ \ \ \ \ \ \ }p=i\sqrt{\frac{\hbar \omega m}{2}}\left( a^{\dagger
}-a\right) ,  \label{2}
\end{equation}%
such that, we can compute any arbitrary expectation values that depend upon
these quantities, merely by knowing the effects of the raising and lowering
operators upon the eigenstates of the Hamiltonian.

From this, we can evaluate that the energy eigenvalues 
\begin{equation}
H^{os}\psi _{n}^{os}(x)=E_{n}\psi _{n}^{os}(x)=\hbar \omega \left( n+\frac{1%
}{2}\right) \psi _{n}^{os}(x);\text{ \ \ \ }n\in 
\mathbb{N}
,  \label{ener}
\end{equation}%
and the normalized condition for the eigenfunctions is verified%
\begin{equation}
\left\langle \psi _{m}^{os}\right. \left\vert \psi _{n}^{os}\right\rangle
=\delta _{mn}.
\end{equation}%
We see that the energy eigenvalues $E_{0}=$\ $\hbar \omega /2$\ \ of the
ground state

\begin{equation}
\psi _{0}^{os}(x)=\frac{1}{\sqrt{2^{n}n!}}\left( \frac{\omega m}{\pi \hbar }%
\right) ^{\frac{1}{4}}\exp \left[ -\frac{\omega m}{2\hbar }x^{2}\right] ,
\label{her}
\end{equation}%
is a very significant physical result because it tells us that the energy of
a system described by a harmonic oscillator potential cannot have zero
energy.

In contrast with the harmonic oscillator, the inverted\ oscillator has a
Hamiltonian with the following form:%
\begin{equation}
H^{r}=\frac{1}{2m}p^{2}-\frac{1}{2}m\omega ^{2}x^{2}=-\frac{\hbar \omega }{2}%
\left( a^{\dagger 2}+a^{2}\right) .  \label{Hinv}
\end{equation}

The Hamiltonian (\ref{Hinv}) is formally obtainable from (\ref{Hosc}) by the
replacement 
\begin{equation}
\omega \rightarrow i\omega ,  \label{iw}
\end{equation}
similarly, the case $(-i\omega )$ would serve equally well.

On the other hand, for an imaginary frequency, i.e. for the inverted
harmonic oscillator, we get

\begin{equation}
a\rightarrow A=e^{i\frac{\pi }{4}}(\sqrt{\frac{m\omega }{2\hbar }}x+\frac{p}{%
\sqrt{2m\omega \hbar }}),  \label{a1}
\end{equation}

\begin{equation}
a^{+}\rightarrow \bar{A}=e^{i\frac{\pi }{4}}(\sqrt{\frac{mw}{2\hbar }}x-%
\frac{p}{\sqrt{2mw\hbar }}),  \label{a+1}
\end{equation}%
\ thus, the Hamiltonian (\ref{Hinv}) can take the following form 
\begin{equation}
H^{r}=\frac{i\hbar \omega }{2}(\bar{A}A+A\bar{A}),  \label{Hr1}
\end{equation}%
where the non-Hermitian ladder operators $\left( A,\bar{A}\right) $ are
characterized by $\left[ A,\bar{A}\right] $ $=1$ in an analogous way to the
ladder operator $\left( a,a^{\dagger }\right) $ for the harmonic oscillator.

Knowing that the eigenfunctions of the harmonic oscillator are normalized,
we ask the question if the inverted oscillator eigenfunctions are also
normalized ? Clearly, they are not $\left\langle \psi _{m}^{r}\right.
\left\vert \psi _{n}^{r}\right\rangle \neq \delta _{mn}.$ This can be seen
when calculating the normalization condition for the ground state $\psi
_{0}^{r}(x)$\ of the obtained inverted oscillator: from Eq. (\ref{her}) by
changing $\omega $ to $i\omega $ 
\begin{equation}
\psi _{0}^{r}(x)=\frac{1}{\sqrt{2^{n}n!}}\left( \frac{i\omega m}{\pi \hbar }%
\right) ^{\frac{1}{4}}\exp \left[ -i\frac{\omega m}{2\hbar }x^{2}\right] .
\end{equation}

One can easily verify that the normalization for this state diverges as
follows: 
\begin{eqnarray}
\left\langle \psi _{0}^{r}\right. \left\vert \psi _{0}^{r}\right\rangle
&=&\int_{-\infty }^{+\infty }\psi _{0}^{\ast r}(x)\psi _{0}^{r}(x)dx  \notag
\\
&=&\frac{1}{2^{n}n!}\left( \frac{\omega m}{\pi \hbar }\right) ^{\frac{1}{2}%
}\int_{-\infty }^{+\infty }dx\rightarrow \infty \text{ \ ,}
\end{eqnarray}

the reason for this divergence is that the substitution $\omega $ by $%
i\omega $ is unsuitable . we will remedy this inconsistency in what follows.

\section{Ladder operators in the inverted harmonic oscillator}

The Hermitian Hamiltonian $H^{r}$ and the non-Hermitian Hamiltonian $%
(-iH^{os})$ are related by a formal replacement $\omega $ $\rightarrow $ $%
i\omega $. The challenge is to establish a consistent relation between the
quantum mechanical formalism for the Hermitian Hamiltonian $H^{r}$and the
non-Hermitian one $(-iH^{os}),$ we propose that instead of considering this
formal transformation, we use the relation that it is valid for any
self-adjoint operator, i.e. observable, in the Hermitian system to possess a
counterpart in the non-Hermitian system given by

\begin{equation}
\rho ^{-1}(-iH^{os})\rho =H^{r}.
\end{equation}

In order to connect the non-Hermitian harmonic oscillator Hamiltonian $%
(-iH^{os})$ to the Hermitian inverted oscillator $H^{r}$, we perform a Dyson
type transformation $\rho $ such that \cite{klimov}

\begin{align}
\rho & =\exp \left\{ -2\left[ \frac{\epsilon }{2}\left( a^{\dagger }a+\frac{1%
}{2}\right) +\mu _{-}\frac{a^{2}}{2}+\mu _{+}\frac{a^{\dagger 2}}{2}\right]
\right\} ,  \notag \\
& =\exp \left[ -\vartheta _{-}\frac{a^{2}}{2}\right] \exp \left[ -\frac{\ln
\vartheta _{0}}{2}\left( a^{\dagger }a+\frac{1}{2}\right) \right] \exp \left[
-\vartheta _{+}\frac{a^{\dagger 2}}{2}\right] ,  \label{D1}
\end{align}%
and 
\begin{align}
\vartheta _{+}& =\frac{2\mu _{+}\sinh \theta }{\theta \cosh \theta -\epsilon
\sinh \theta },  \notag \\
\vartheta _{0}& =\left( \cosh \theta -\frac{\epsilon }{\theta }\sinh \theta
\right) ^{-2}=\mu _{+}\mu _{-}-\chi ,  \notag \\
\vartheta _{-}& =\frac{2\mu _{-}\sinh \theta }{\theta \cosh \theta -\epsilon
\sinh \theta }, \\
\chi & =-\frac{\cosh \theta +\frac{\epsilon }{\theta }\sinh \theta }{\cosh
\theta -\frac{\epsilon }{\theta }\sinh \theta }\text{ \ \ \ \ ,\ \ \ \ \ }%
\theta =\sqrt{\epsilon ^{2}-4\mu _{+}\mu _{-}},  \notag
\end{align}%
where $\epsilon $ is a real parameter whereas $\mu _{+}$ and $\mu _{-}$ are
complex ones.

With the help of the following relations\ 
\begin{equation}
\left\{ 
\begin{array}{c}
\exp \left[ \vartheta _{-}\frac{a^{2}}{2}\right] \left( a^{\dagger }a+\frac{1%
}{2}\right) \exp \left[ -\vartheta _{-}\frac{a^{2}}{2}\right] =\left(
a^{\dagger }a+\frac{1}{2}\right) +\vartheta _{-}a^{2} \\ 
\exp \left[ \vartheta _{+}\frac{a^{\dagger 2}}{2}\right] \left( a^{\dagger
}a+\frac{1}{2}\right) \exp \left[ -\vartheta _{+}\frac{a^{\dagger 2}}{2}%
\right] =\left( a^{\dagger }a+\frac{1}{2}\right) -\vartheta _{+}a^{\dagger 2}%
\end{array}%
\right. ,
\end{equation}%
\begin{equation}
\left\{ 
\begin{array}{c}
\exp \left[ \frac{\ln \vartheta _{0}}{2}\left( a^{\dagger }a+\frac{1}{2}%
\right) \right] a^{2}\exp \left[ -\frac{\ln \vartheta _{0}}{2}\left(
a^{\dagger }a+\frac{1}{2}\right) \right] =\frac{a^{2}}{\vartheta _{0}} \\ 
\exp \left[ \vartheta _{+}\frac{a^{\dagger 2}}{2}\right] a^{2}\exp \left[
-\vartheta _{+}\frac{a^{\dagger 2}}{2}\right] =a^{2}-2\vartheta _{+}\left(
a^{\dagger }a+\frac{1}{2}\right) +\vartheta _{+}^{2}a^{\dagger 2}%
\end{array}%
\right. ,
\end{equation}%
\begin{equation}
\left\{ 
\begin{array}{c}
\exp \left[ \frac{\ln \vartheta _{0}}{2}\left( a^{\dagger }a+\frac{1}{2}%
\right) \right] a^{\dagger 2}\exp \left[ -\frac{\ln \vartheta _{0}}{2}\left(
a^{\dagger }a+\frac{1}{2}\right) \right] =\vartheta _{0}a^{\dagger 2} \\ 
\exp \left[ \vartheta _{-}\frac{a^{2}}{2}\right] a^{\dagger 2}\exp \left[
-\vartheta _{-}\frac{a^{2}}{2}\right] =a^{\dagger 2}+2\vartheta _{-}\left(
a^{\dagger }a+\frac{1}{2}\right) +\vartheta _{-}^{2}a^{2}%
\end{array}%
\right. ,
\end{equation}%
we deduce, under the action of the operator $\rho $, the transformed
Hamiltonian of the harmonic oscillator : 
\begin{eqnarray}
\rho ^{-1}H^{os}\rho &=&\hbar \omega \rho ^{-1}\left( a^{\dagger }a+\frac{1}{%
2}\right) \rho ,  \notag \\
&=&\frac{\hbar \omega }{\vartheta _{0}}\left\{ \left[ \vartheta
_{0}-2\vartheta _{+}\vartheta _{-}\right] \left( a^{\dagger }a+\frac{1}{2}%
\right) +\left[ \vartheta _{-}\vartheta _{+}^{2}-\vartheta _{0}\vartheta _{+}%
\right] a^{\dagger 2}+\vartheta _{-}a^{2}\right\} .  \label{h1}
\end{eqnarray}

We notice that Eq.(\ref{h1}) and Eq.(\ref{Hinv}) have the same structure in
their operator content provided that we impose on the parameters $\left(
\vartheta _{+},\vartheta _{-},\text{ }\vartheta _{0}\right) $ the following
conditions 
\begin{equation}
\vartheta _{+}=-i,\text{ }\vartheta _{-}=\frac{i}{2}\text{, }\vartheta
_{0}=1,  \label{3}
\end{equation}%
from these constraints, the Dyson operator Eq.(\ref{D1}) takes now the
simplified form\footnote{%
It is useful to note that the following simplified transformations (\ref{D2}%
) has been introduced in Ref. \cite{Ken} as footnote.} 
\begin{eqnarray}
\rho &=&\exp [-\frac{i}{4}a^{2}]\exp [\frac{i}{2}a^{\dagger 2}],  \notag \\
\rho ^{-1} &=&\exp [-\frac{i}{2}a^{\dagger 2}]\exp [\frac{i}{4}a^{2}],
\label{D2}
\end{eqnarray}%
it follows that the two Hamiltonians $H^{os}$ and $H^{r}$ are allied to each
other as 
\begin{equation}
\rho ^{-1}H^{os}\rho =i\frac{\hbar \omega }{2}\left( a^{\dagger
2}+a^{2}\right) =iH^{r}.  \label{Hr2}
\end{equation}

One can verify that in the case of the inverted oscillator, the form of
Hamiltonian in the last equation looks like 
\begin{equation}
H^{r}=\frac{i\hbar \omega }{2}(\bar{A}A+A\bar{A}),  \label{Hr3}
\end{equation}%
where the ladder operators $\left( A,\bar{A}\right) $ are linked to the
ladder operators (\ref{L1}) through the transformation 
\begin{equation}
A=\rho ^{-1}a\rho =a+ia^{\dagger },  \label{A1}
\end{equation}%
\begin{equation}
\bar{A}=\rho ^{-1}a^{\dagger }\rho =\frac{1}{2}\left( a^{\dagger }+ia\right)
,  \label{A2}
\end{equation}%
and satisfy the following commutation relation $\left[ A,\bar{A}\right] $ $%
=1.$Then, we can deduce that their Fock eigenstates $\left\vert
n^{r}\right\rangle $ are related to $\left\vert n^{os}\right\rangle $ by the
invertible operator $\rho $ as 
\begin{equation}
\left\vert n^{r}\right\rangle =\rho ^{-1}\left\vert n^{os}\right\rangle .
\label{r}
\end{equation}

For instance, the pseudo-Hermitian quadratures $(X,P)$ corresponding in the
Hermitian system to the coordinate and momentum operators $(x,p)$ (see Eqs. (%
\ref{2})) respectively, are now 
\begin{eqnarray}
X &=&\rho ^{-1}x\rho =\sqrt{\frac{\hbar }{2\omega m}}\rho ^{-1}\left(
a^{\dagger }+a\right) \rho \text{ }  \notag \\
&=&\sqrt{\frac{\hbar }{2\omega m}}\left( A+\bar{A}\right) ,  \label{X1}
\end{eqnarray}%
\begin{eqnarray}
\ P &=&\rho ^{-1}p\rho =i\sqrt{\frac{\hbar \omega m}{2}}\rho ^{-1}\left(
a^{\dagger }-a\right) \rho  \notag \\
&=&i\sqrt{\frac{\hbar \omega m}{2}}\left( \bar{A}-A\right) .  \label{X2}
\end{eqnarray}

Knowing that any observable $o$ in the Hermitian system possesses a
counterpart $O$ in the pseudo-Hermitian system given by

\begin{equation}
O=\rho ^{-1}o\rho ,  \label{O}
\end{equation}%
one can deduce the useful representation of $\left( A,\bar{A}\right) $ in
terms of $(X,P)$ as 
\begin{equation}
A=\sqrt{\frac{m\omega }{2\hbar }}X+i\frac{1}{\sqrt{2m\hbar \omega }}P,
\label{A1'}
\end{equation}%
\begin{equation}
\bar{A}=\sqrt{\frac{m\omega }{2\hbar }}X-i\frac{1}{\sqrt{2m\hbar \omega }}P.
\label{A2'}
\end{equation}

Thereby, the Hamiltonian (\ref{Hr3}) can be written in terms of $X$ and $P$
as 
\begin{equation}
H^{r}=\frac{i}{2}(\frac{P^{2}}{m}+m\omega ^{2}X^{2}).  \label{Hr3'}
\end{equation}

This leads to the equations of motion of the\ inverted oscillator. Indeed,
using the Heisenberg equations of motion and $\left[ X,P\right] =i\hbar $,
we have for $X$ and $P$: 
\begin{eqnarray}
\frac{dX}{dt} &=&\frac{1}{i\hbar }\left[ X,\frac{i}{2}(\frac{P^{2}}{m}%
+m\omega ^{2}X^{2})\right] =i\frac{P}{m}.  \notag \\
\frac{dP}{dt} &=&\frac{1}{i\hbar }\left[ P,\frac{i}{2}(\frac{P^{2}}{m}%
+m\omega ^{2}X^{2})\right] =-im\omega ^{2}X.  \label{Hr4}
\end{eqnarray}

Taking another time derivative of $dX/dt$,\ we get the usual equation of
motion for the inverted oscillator%
\begin{equation}
\frac{d^{2}X}{dt^{2}}-\omega ^{2}X=0,  \label{Hr5}
\end{equation}

\section{Coherent states for the inverted oscillator}

The best way to present the inverted coherent states is by translating their
definitions into the language of the coherent states of\ the harmonic
oscillator which are summarized in what follows. Coherent states, or
semi-classic states, are remarkable quantum states that were originally
introduced in 1926 by Schr\"{o}dinger for the Harmonic oscillator \cite%
{Schrodinger} where the mean values of the position and momentum operators
in these states have properties close to the classical values of the
position $x_{c}(t)$ and the momentum $p_{c}(t).$ In particular, the coherent
states of the harmonic oscillator$\left\vert \alpha ^{os}\right\rangle $ 
\cite{Glauber}-\cite{Sudarshan} may be obtained in\ different but equivalent
ways:

(i) as eigenstates of the annihilation operator; 
\begin{equation}
a\left\vert \alpha \right\rangle ^{os}=\alpha \left\vert \alpha
\right\rangle ^{os},  \label{c1}
\end{equation}%
with eigenvalues $\alpha \in $ $C.$

(ii) as a displacement of the vacuum $\left\vert 0\right\rangle ^{^{os}}$,
where the displacement operator 
\begin{equation}
D^{os}\left( \alpha \right) =\exp [\alpha ^{\ast }a^{\dagger }-\alpha a],
\label{c2}
\end{equation}%
can be used to generate the coherent state 
\begin{equation}
\left\vert \alpha \right\rangle ^{os}=D^{os}\left( \alpha \right) \left\vert
0\right\rangle ^{os},  \label{c3}
\end{equation}%
\ \ \ \ (iii) as states\ that minimize the Heisenberg uncertainty principle 
\begin{equation}
\text{\ \ }\Delta x\Delta p=\frac{\hbar }{2}.
\end{equation}

Coherent states form an over-complete set of states. The identity operator $%
I $ is written in terms of coherent states as 
\begin{equation}
\frac{1}{\pi }\int \left\vert \alpha \right\rangle ^{os\text{ \ }%
os}\left\langle \alpha \right\vert d^{2}\alpha =I\mathbf{.}  \label{c4}
\end{equation}%
The solution for the harmonic oscillator Hamiltonian for an initial coherent
state is given in the following simple form%
\begin{equation}
\left\vert \alpha ,t\right\rangle ^{os}=e^{-i\frac{\omega t}{2}}\left\vert
\alpha e^{-i\omega t}\right\rangle ^{os},  \label{c5}
\end{equation}%
i.e., a coherent state that rotates with the harmonic oscillator frequency.

In analogy with the usual coherent states, we use the annihilation $A=\rho
^{-1}a\rho $ and creation $\bar{A}=\rho ^{-1}a^{\dagger }\rho $ operators
which are very convenient to study the inverted coherent states. We
emphasize the use of the metric $\eta =\rho ^{\dagger }\rho $ operator such
as $\left( iH^{os}\right) ^{\dagger }=\eta \left( iH^{os}\right) \eta ^{-1}$%
, i.e. $\left( iH^{os}\right) $ is $\eta $-pseudo-Hermitian with respect to
a positive-definite inner product defined by $\left\langle .,.\right\rangle
_{\eta }=\left\langle .|\eta |.\right\rangle :$ 
\begin{equation}
^{r}\left\langle n\right\vert \eta \left\vert m\right\rangle ^{r}=\text{ }%
^{os}\left\langle n\right. \left\vert m\right\rangle ^{os}=\delta _{mn},
\end{equation}%
which indicates that the Fock states are linked to each other as 
\begin{equation}
\left\vert n\right\rangle ^{r}=\rho ^{-1}\left\vert n\right\rangle ^{os},
\label{nR}
\end{equation}%
additionally, the vacuum state of the inverted oscillator $\left\vert
0\right\rangle ^{r}$ ($A\left\vert 0\right\rangle ^{r}=0$) and the vacuum
state of the\ harmonic oscillator $\left\vert 0\right\rangle ^{os}$ are
related as $\left\vert 0\right\rangle ^{r}=\rho ^{-1}\left\vert
0\right\rangle ^{os}.$

The coherent states for the inverted harmonic oscillator are defined as
eigenstates of the corresponding annihilation operator $A$ 
\begin{equation}
A\left\vert \alpha \right\rangle ^{r}=\alpha \left\vert \alpha \right\rangle
^{r},\text{ \ \ \ }\alpha \in \mathbb{C}.  \label{a4}
\end{equation}

with\ 
\begin{equation}
\left\vert \alpha \right\rangle ^{r}=\rho ^{-1}\left\vert \alpha
\right\rangle ^{os}.  \label{3.2.1.5}
\end{equation}

Particularly, the normalization condition%
\begin{equation}
^{os}\left\langle \alpha \right. \left\vert \alpha \right\rangle ^{os}=1,
\end{equation}%
leads to%
\begin{equation}
^{r}\left\langle \alpha \right\vert \eta \left\vert \alpha \right\rangle
^{r}=1,
\end{equation}%
and then the integral 
\begin{equation}
\frac{1}{\pi }\int_{\mathbb{C}}\rho \left\vert \alpha \right\rangle ^{r}%
\text{ }^{r}\left\langle \alpha \right\vert \rho ^{+}d\alpha ^{\ast }d\alpha
=I,
\end{equation}%
is an identity operator.

These inverted coherent states $\left\vert \alpha \right\rangle ^{r}$ can
also be generated respectively from the vacuum states $\left\vert
0\right\rangle ^{r}$ by the action of displacement operator $D^{r}(\alpha )$
,

\begin{equation}
\left\vert \alpha \right\rangle ^{r}=D^{r}\left( \alpha \right) \left\vert
0\right\rangle ^{r}=\exp \left[ \alpha \overline{A}-\alpha ^{\ast }A\right]
\left\vert 0\right\rangle ^{r},
\end{equation}%
we note that $D^{r}\left( \alpha \right) $\ is related to $D^{os}\left(
\alpha \right) $ as%
\begin{equation}
D^{r}\left( \alpha \right) =\rho ^{-1}D^{os}\left( \alpha \right) \rho .
\end{equation}

Using the Hamiltonian (\ref{Hr3}), we deduce the evolution of an initial
inverted coherent state in the following simple form%
\begin{equation}
\left\vert \alpha ,t\right\rangle ^{r}=e^{-\frac{i}{\hbar }H^{r}t}\left\vert
\alpha \right\rangle ^{r}=e^{-\frac{\left\vert \alpha \right\vert ^{2}}{2}%
}e^{\frac{^{\omega t}}{2}}e^{\omega \bar{A}At}\tsum_{n}\frac{(\alpha )^{n}}{%
\sqrt{n!}}\left\vert n\right\rangle ^{r}.
\end{equation}

Introducing $e^{\omega \bar{A}At}$ into the sum, and using the fact that the
states $\left\vert n\right\rangle ^{r}$ are eigenstates of the number
operator$\bar{A}A$, we have

\begin{equation}
\left\vert \alpha ,t\right\rangle ^{r}=e^{-\frac{\left\vert \alpha e^{\omega
t}\right\vert ^{2}}{2}}\tsum_{n}\frac{(\alpha e^{\omega t})^{n}}{\sqrt{n!}}%
\left\vert n\right\rangle ^{r}=e^{\frac{\omega t}{2}}\left\vert \alpha
e^{\omega t}\right\rangle ^{r}.  \label{co}
\end{equation}

Since our aim is to compute the Heisenberg uncertainty relations in the
position and the momentum, it is required to calculate the expectation\
values of the canonical variables and their squares in the inverted coherent
states. Then, by using the relation (\ref{O}) in the non-Hermitian system,
the expectation value of an arbitrary operator $O=X,X^{2},P$ and $P^{2}$ can
be evaluated from 
\begin{eqnarray}
\left\langle O\right\rangle _{\eta }\text{ } &=&\text{ }^{r}\left\langle
\alpha ,t\right\vert \eta O\left\vert \alpha ,t\right\rangle ^{r}=\text{ }%
^{r}\left\langle \alpha ,t\right\vert \rho ^{+}o\rho \text{ }\left\vert
\alpha ,t\right\rangle ^{r}  \notag \\
&=&e^{-\frac{\left\vert \alpha e^{\omega t}\right\vert ^{2}}{2}%
}\tsum_{n}\tsum_{m}\frac{(\alpha ^{\ast }e^{\omega t})^{m}}{\sqrt{m!}}\frac{%
(\alpha e^{\omega t})^{n}}{\sqrt{n!}}\text{ }^{os}\left\langle m\right\vert
o\left\vert n\right\rangle ^{os}.  \label{inc}
\end{eqnarray}

Using the above equation, the expectation values of $X$ and $P$ in the state 
$\left\vert \alpha ,t\right\rangle ^{r}$ are easily evaluated:

\begin{equation}
\left\langle X\right\rangle _{\eta }\text{ }=\text{ }e^{-\left\vert \alpha
e^{\omega t}\right\vert ^{2}}\tsum_{n}\tsum_{m}\frac{(\alpha ^{\ast
}e^{\omega t})^{m}}{\sqrt{m!}}\frac{(\alpha e^{\omega t})^{n}}{\sqrt{n!}}%
\sqrt{\frac{\hbar }{2\omega m}}\text{ }^{os}\left\langle m\right\vert \left(
a^{\dagger }+a\right) \left\vert n\right\rangle ^{os}=\sqrt{\frac{\hbar }{%
2m\omega }}\left[ \alpha +\alpha ^{\ast }\right] e^{\omega t},  \label{X}
\end{equation}

\begin{equation}
\left\langle P\right\rangle _{\eta }\text{ }=\text{ }e^{-\left\vert \alpha
e^{\omega t}\right\vert ^{2}}\tsum_{n}\tsum_{m}\frac{(\alpha ^{\ast
}e^{\omega t})^{m}}{\sqrt{m!}}\frac{(\alpha e^{\omega t})^{n}}{\sqrt{n!}}i%
\sqrt{\frac{\hbar \omega m}{2}}\text{ }^{os}\left\langle m\right\vert \left(
a^{\dagger }-a\right) \left\vert n\right\rangle ^{os}=-i\sqrt{\frac{m\omega
\hbar }{2}}\left[ \alpha -\alpha ^{\ast }\right] e^{\omega t},  \label{P}
\end{equation}%
and follow classical physics; i.e.

\begin{equation}
\left\langle X\right\rangle _{\eta }=x_{c},\ \ \ \ \left\langle
P\right\rangle _{\eta }=p_{c},
\end{equation}%
where the subscript $c$ indicate classical. This is why we call these
inverted coherent states "quasi-classical states".

Let us now evaluate the uncertainty in the position and the momentum.

\begin{equation}
\left\langle X^{2}\right\rangle _{\eta }=\left\langle \alpha ,t\right\vert
\eta X^{2}\left\vert \alpha ,t\right\rangle ^{r}=\frac{\hbar }{2m\omega }%
\left[ \alpha ^{2}e^{2\omega t}+\alpha ^{\ast 2}e^{2\omega t}+2(\left\vert
\alpha \right\vert ^{2}e^{2\omega t}+\frac{1}{2})\right] ,  \label{X 2}
\end{equation}

\begin{equation}
\left\langle P^{2}\right\rangle _{\eta }=\left\langle \alpha ,t\right\vert
\eta P^{2}\left\vert \alpha ,t\right\rangle ^{r}=\frac{-im\omega \hbar }{2}%
\left[ \alpha ^{2}e^{2\omega t}+\alpha ^{\ast 2}e^{2\omega t}-2(\left\vert
\alpha \right\vert ^{2}e^{2\omega t}+\frac{1}{2})\right] .  \label{P2}
\end{equation}

It is well known that the position uncertainty can be derived from $\Delta X=%
\sqrt{\left\langle X^{2}\right\rangle _{\eta }-\left\langle X\right\rangle
_{\eta }^{2}}$. Then using (\ref{X}) and (\ref{X 2}), we have 
\begin{equation*}
\Delta X=\sqrt{\frac{\hbar }{2m\omega }}.
\end{equation*}

Similarly, from Eqs. (\ref{P}) and (\ref{P2}), we also have the momentum
uncertainty such that

\begin{equation*}
\Delta P=\sqrt{\frac{m\omega \hbar }{2}.}
\end{equation*}

Thus, the uncertainty product for canonical variables $X$ and $P$ is given
by 
\begin{equation*}
\Delta X\Delta P=\frac{\hbar }{2}.
\end{equation*}

Therefore, the inverted coherent states are a minimum-uncertainty states and
the time evolution of an initially inverted coherent state can be regarded
as the quantum analog of a classical trajectory.

\section{Conclusion}

\ We have briefly summarized\ in section 2, some properties of the standard
harmonic and inverted oscillators.

We have proposed a scheme that permits relating a regular harmonic
oscillator to an inverted oscillator by using a time-independent Dyson
metric which allowed us to introduce the annihilation $A=\rho ^{-1}a\rho $\
and creation $\bar{A}=\rho ^{-1}a^{\dagger }\rho $ operators associated to
the inverted harmonic oscillator. These operators are the basis of the
definition of coherent states for inverted oscillator and their
corresponding eigenstates and eigenvalues. Once the Dyson operator has been
introduced, and therefore the metric operators, it is straightforward to
extend these considerations to the associated eigenstates and inner product
structures on the physical Hilbert space.

Coherent states of the inverted harmonic oscillator are constructed in
different forms :

(1) as eigenstates of the annihilation operator $A$;

(2) as a displacement of the inverted vacuum $\exp \left[ \alpha \overline{A}%
-\alpha ^{\ast }A\right] \left\vert 0\right\rangle ^{r}$,

(3) as states whose averages follow the classical trajectories of $X$, $P$
and $H^{r}.$

However, the coherent states for the inverted oscillator constitute "minimum
uncertainty" wave packets. Therefore, the time evolution of an initially
coherent state can be regarded as the quantum analog of a classical
trajectory.

\paragraph{ACKNOWLEDGMENTS}

The authors are grateful to Dr N. Mana and Dr K. Benzid for the valuable
suggestions that led to considerable improvement in the presentation of this
paper.

\end{document}